\documentstyle[aasms4]{article}

\begin{document}
\title{RELATIVISTIC THERMAL BREMSSTRAHLUNG GAUNT 
FACTOR FOR THE INTRACLUSTER PLASMA. III. ANALYTIC FITTING FORMULA FOR THE
NONRELATIVISTIC EXACT GAUNT FACTOR}

\author{NAOKI ITOH, SHUGO KUSANO, AND TSUYOSHI SAKAMOTO,} 
\affil{Department of Physics, Sophia University, 7-1 Kioi-cho, Chiyoda-ku, Tokyo, 102-8554, Japan;}
\affil{n\_itoh, s-kusano, t-sakamo@hoffman.cc.sophia.ac.jp}

\author{SATOSHI NOZAWA}

\affil{Josai Junior College for Women, 1-1 Keyakidai, Sakado-shi, Saitama, 350-0290, Japan;}

\affil{snozawa@galaxy.josai.ac.jp}

\centerline{AND}

\author{YASUHARU KOHYAMA}

\affil{Fuji Research Institute Corporation, 2-3 Kanda-Nishiki-cho, Chiyoda-ku, Tokyo, 101-8443, Japan;}
\affil{kohyama@star.fuji-ric.co.jp}

\begin{abstract}

  We present an accurate analytic fitting formula for the thermal bremsstrahlung Gaunt factor in the nonrelativistic limit. The fitting formula excellently reproduces the numerical results of the calculation carried out by the present authors using the method of Karzas and Latter. The present analytic fitting formula will be useful for the analysis of the X-ray emission which comes from the intracluster plasmas with relatively low temperatures as well as the other X-ray sources. 

\end{abstract}


\section{INTRODUCTION}

  The present authors have recently carried out accurate calculations on the relativistic thermal bremsstrahlung Gaunt factor for the intracluster plasma (Nozawa, Itoh, \& Kohyama 1998). They have also presented accurate analytic fitting formulae which summarize the numerical resluts of the calculations (Itoh et al. 1999). Their calculation is based on the method of Itoh and his collaborators (Itoh, Nakagawa, \& Kohyama 1985; Nakagawa, Kohyama, \& Itoh 1987; Itoh, Kojo, \& Nakagawa 1990; Itoh et al. 1991, 1997). 
  In calculating the relativistic thermal bremsstrahlung Gaunt factor for the high-temperature, low-density plasma, Nozawa, Itoh, \& Kohyama (1998) have made use of the Bethe-Heitler cross section (Bethe \& Heitler 1934) corrected by the Elwert factor (Elwert 1939). They have also calculated the Gaunt factor by using the Coulomb-distorted wave functions for nonrelativistic electrons following the method of Karzas \& Latter (1961). 
  In Itoh et al. (1999), the present authors have constructed accurate analytic fitting formulae by combining the relativistic Elwert Gaunt factor with the nonrelativistic exact Gaunt factor. The former Gaunt factor is accurate at high temperatures, whereas the latter is accurate at low temperatures. For the plasmas with relatively low temperatures, the nonrelativistic exact Gaunt factor alone is sufficient for the analysis of the radiation which comes from these plasmas. Therefore, it is worthwhile to present an accurate analytic fitting formula which reproduces the numerical results of the calculations for the nonrelativistic exact Gaunt factor which have been reported in Nozawa, Itoh, \& Kohyama (1998).
  The present paper is organized as follows. We will present the accurate analytic fitting formula in $\S$2. Concluding remarks will be given in $\S$3.

\section{ANALYTIC FITTING FORMULA}
  The thermal bremsstrahlung emissivity in the nonrelativistic limit is expressed in terms of the nonrelativistic exact Gaunt factor $g_{\rm NR}$ (Nozawa, Itoh, \& Kohyama 1998) by 

\begin{eqnarray}
< W(\omega) >_{\rm NR} d \omega & = & 1.426 \times 10^{-27}g_{\rm NR}( \gamma^2 ,u)
\left[ n_{e}({\rm cm^{-3}}) \right] \left[n_{j}({\rm cm^{-3}}) \right] Z_{j}^{2} \left[ T({\rm K}) \right]^{1/2} \nonumber  \\
& \times & e^{-u} \ du \hspace{0cm} \ {\rm ergs\,\, s^{-1} \, cm^{-3}} \, , \\
u        & \equiv & \frac{\hbar \omega}{k_{B}T}  \, , \\
\gamma^2 & \equiv & \frac{{Z_j}^2{\rm Ry}}{k_B T} = {Z_j}^2 \frac{1.579 \times 10^5{\rm K}}{T} \, .
\end{eqnarray}

  In the above, $\omega$ is the angular frequency of the emitted photon, $T$ is the temperature of the electrons, $n_e$ is the number density of the electrons, $n_j$ is the number density of the ions with the charge $Z_j$.
  It should be noted that the thermal bremsstrahlung emissivity in the nonrelativistic limit is a function of $\gamma^2$ and $u$ only. It does not depend on $Z_j$ and $T$ separately, but on the ratio ${Z_j}^2 / T $. This is a remarkable fact for nonrelativistic electrons.

  In Figure 1 we show the nonrelativistic Gaunt factor as a function of $u$ for various values of $\gamma^2$. In Figure 2 we show the nonrelativistic Gaunt factor as a function of $\gamma^2$ for various values of $u$.

  We give an analytic fitting formula for the nonrelativistic exact Gaunt factor. The range of the fitting is $-$3.0 $\leq$ $\log_{10} \gamma^2 \leq$ 2.0, $-$4.0 $\leq$ $\log_{10} u$ $\leq$ 1.0. We express the Gaunt factor by 
\begin{eqnarray}
g_{\rm NR} & =      &  \sum^{10}_{i,j = 0} b_{ij}\Gamma^i U^j  \,  ,  \\
\Gamma & \equiv & \frac{1}{2.5}[\log_{10} \gamma^2 + 0.5] \,  ,  \\
U      & \equiv & \frac{1}{2.5}[\log_{10} u + 1.5] .
\end{eqnarray}
  The coefficients $b_{ij}$ are presented in TABLE 1.
The accuracy of the fitting is generally better than 0.1\%.

\section{CONCLUDING REMARKS}

  We have presented accurate analytic fitting formulae for the nonrelativistic exact Gaunt factor for thermal bremsstrahlung. The analytic fitting formulae have been constructed to reproduce the numerical results of the calculation by the method of Karzas \& Latter (1961) reported in Nozawa, Itoh, \& Kohyama (1998). The accuracy of the fitting is generally better than 0.1\%. The present fitting formula can be used widely as far as the electrons are nonrelativistic. 

We thank Professor Y. Oyanagi for allowing us to use the least square
fitting program SALS.  This work is financially supported in part
by the Grant-in-Aid of Japanese Ministry of Education, Science, Sports,
and Culture under the contract \#10640289.

\newpage
\begin{center}
{\bf Figure Legends} \\
\end{center}
\begin{tabbing}
FIG.1. \= Nonrelativistic exact Gaunt factor as a function of $u$ for various values of $\gamma^2$. \\
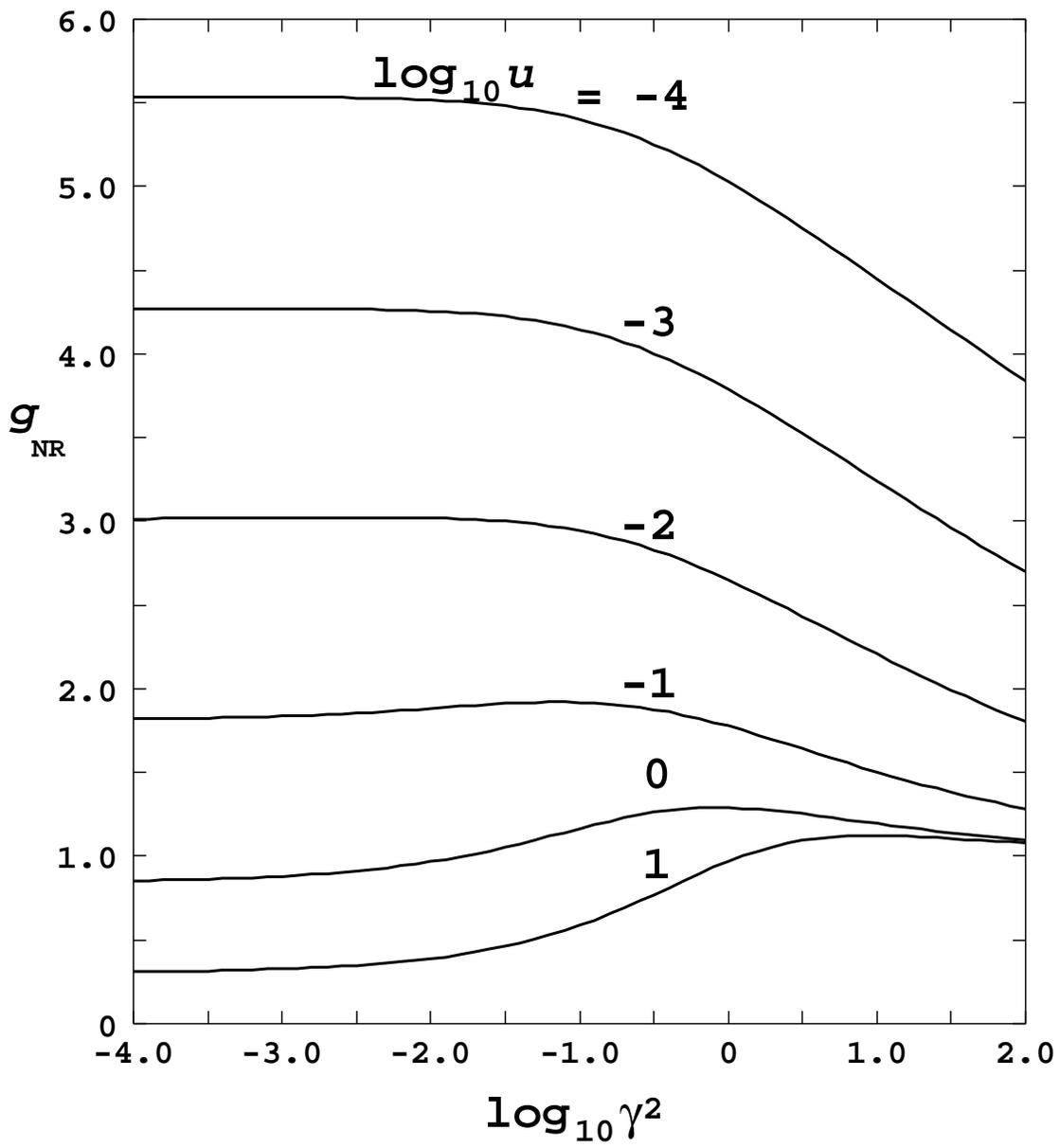
FIG.2. \= Nonrelativistic exact Gaunt factor as a function of $\gamma^2$ for various values of $u$.
\end{tabbing}
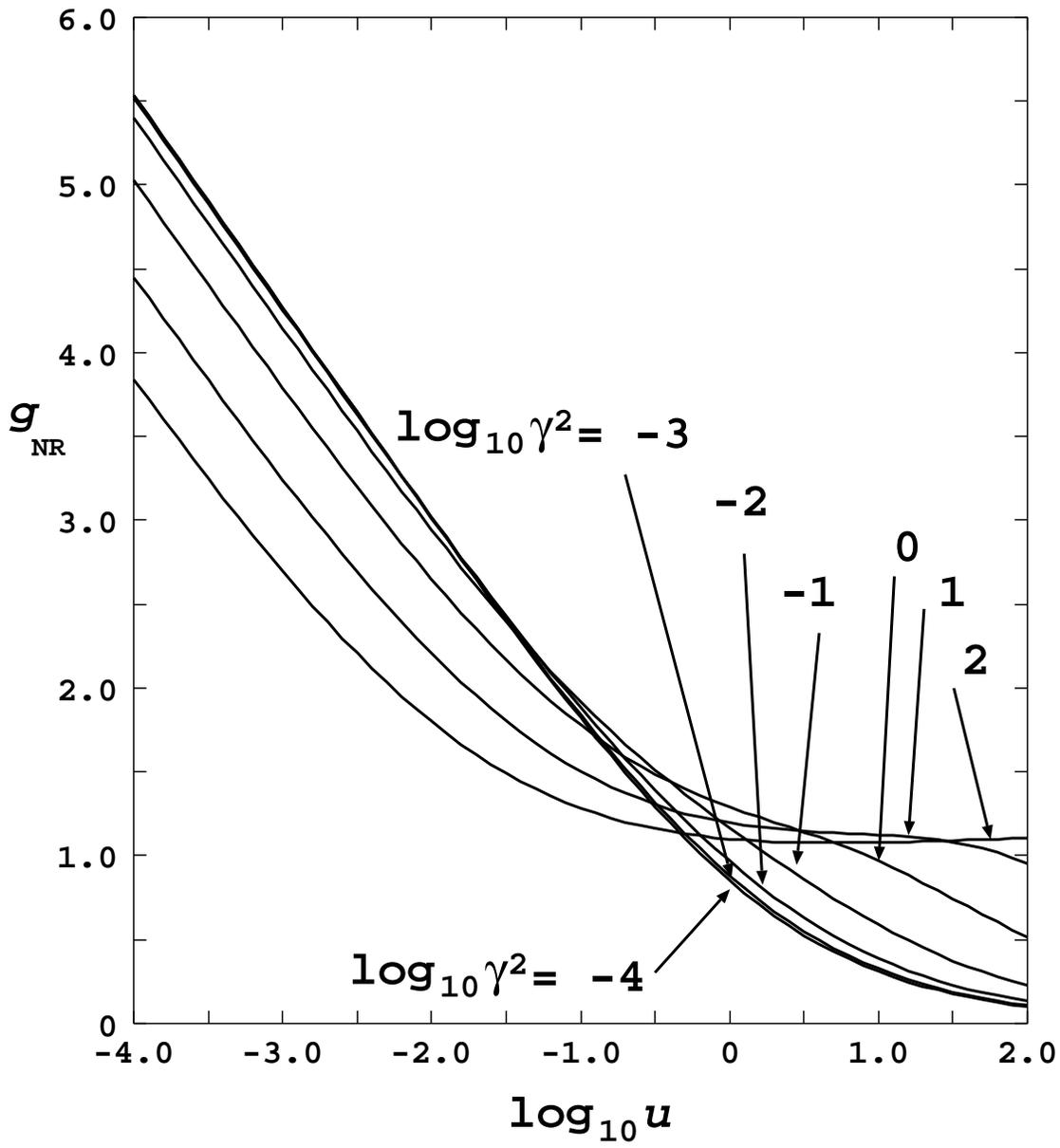
\end{document}